# Nano-Patterned Pt-Based Metallic Glass Electrocatalysts with In-Situ Copper Oxide Foam for Enhanced Hydrogen Evolution

Fei-Fan Cai[*,1,2], Baran Sarac[2], Adnan Akman[3], Juan J. Londoño[1,4], Selin Gümrükcü[2,5], Lukas Schweiger[1], Martin Hantusch[3], Jan Schroers[6], Andreas Blatter[4], Annett Gebert[3], Florian Spieckermann[*,1], Jürgen Eckert[1,2]

[1] Department of Materials Science, Chair of Materials Physics, Montanuniversität Leoben, Jahnstraße 12, A-8700 Leoben, Austria, f.cai@stud.unileoben.ac.at, juan-jose.londono-rueda@stud.unileoben.ac.at, lukas.schweiger@unileoben.ac.at, florian.spieckermann@unileoben.ac.at, juergen.eckert@unileoben.ac.at

[2] Erich Schmid Institute of Materials Science, Austrian Academy of Sciences, Jahnstraße 12, A-8700 Leoben, Austria, fei-fan.cai@oeaw.ac.at, celinegumrukcu@gmail.com, baransarac@gmail.com, juergen.eckert@oeaw.ac.at

[3] Leibniz Institute for Solid State and Materials Research, Helmholtzstr. 20, D-01069 Dresden, Germany, a.akman@ifw-dresden.de, a.gebert@ifw-dresden.de, m.hantusch@ifw-dresden.de

[4] Research and Development Department, PX Services, 2300 La Chaux-de-Fonds, Switzerland, andreas.blatter@pxgroup.com

[5] Department of Chemistry, Istanbul Technical University, 34469 Istanbul, Türkiye, gumrukcuse@itu.edu.tr

[6] Department of Mechanical Engineering and Materials Science, Yale University, New Haven, CT 06511, USA, jan.schroers@yale.edu

**Corresponding author**

Florian Spieckermann, florian.spieckermann@unileoben.ac.at, Department of Materials Science, Chair of Materials Physics, Montanuniversität Leoben, Jahnstraße 12, A-8700 Leoben, Austria

Fei-Fan Cai, f.cai@stud.unileoben.ac.at, Erich Schmid Institute of Materials Science, Austrian Academy of Sciences, Jahnstraße 12, A-8700 Leoben, Austria

## Keywords

**Abstract**

Hydrogen is a promising energy carrier for replacing fossil fuels, and hydrogen production via hydrogen evolution reaction (HER) is an environmentally friendly option if electrocatalysts with low overpotentials and long-term stability are used. In this work, the electrocatalytic performance of $Pt_{57.5}Cu_{14.7}Ni_{5.3}P_{22.5}$ bulk metallic glass (BMG) with flat, micro-patterned, and nano-patterned surfaces for HER in 0.5 M $H_2SO_4$ is studied. The nano-patterned Pt-BMG demonstrates outstanding long-term stability and self-improving behavior with a final overpotential of 150 mV and a Tafel slope of 42 mV dec$^{-1}$ after 1000 linear sweep voltammetry (LSV) cycles, which is respectively 42% and 37% lower than in the first LSV cycle. X-ray photoelectron spectroscopy (XPS) and Auger electron spectroscopy (AES) indicate the formation of a layer of $CuO/Cu_2O$ foam deposited on top of the nano-patterned surface during the stability test of 1000 LSV cycles. A three-step process is proposed to explain the formation of $Cu_xO$ foam via dynamic hydrogen bubble templating (DHBT) electrodeposition from Cu dissolution of the Pt-BMG without using copper salt. This work provides a method to create $Cu_xO$ foams that could be used for various applications. Moreover, nano-patterned BMGs with DHBT deposition offer a feasible strategy to synthesize metal or metal-oxide foams.

## 1. Introduction

Energy is an inseparable part of modern life, and it is required to perform virtually all human activities, from personal devices to the socio-economic development of a country.[1–3] Hydrogen ($H_2$) emerges as a viable energy carrier for both storage and transportation due to its notable benefits, including high energy density ( 140 MJ/kg), versatile applicability, and zero carbon emissions if produced by renewable energy sources.[1,3–6] For efficient and sustainable hydrogen production, the hydrogen evolution reaction (HER) from water electrolysis is the most feasible option.[7–11] However, a highly active and long-lasting catalyst is essential to accelerate the production of hydrogen gas by HER.[12] Pt-based materials have been proven to be the most efficient catalysts for HER in acidic electrolytes, yet their high cost, poor stability, and scarcity severely limit their wide spread utilization.[8,13–15]

Dynamic hydrogen bubble templating (DHBT) electrodeposition is a technique that uses the evolving hydrogen bubbles in HER as a dynamic template to synthesize porous metal foams.[16,17] During DHBT electrodeposition, metal electrodeposition and hydrogen evolution happen simultaneously at a cathode with a high overpotential. Consequently, hydrogen bubbles act as a dynamic template, obstructing metal ion–cathode contact, so that metal ions can only be deposited between bubbles, creating porous metal structures.[16,17]

Bulk metallic glasses (BMGs) have unique mechanical, physical, and chemical properties compared to crystalline alloys, thanks to their disordered atomic structure.[12,18–20] BMGs can be shaped and patterned via thermoplastic forming (TPF) which is based on viscous flow in their supercool liquid region (SCLR).[21–25] Noble metal-based BMGs, such as Pt- and Pd-based BMGs, have been used in jewelry and implant applications.[18,21] For HER applications, BMGs could be a strategy to alleviate the cost issues of noble elements in the catalysts by forming alloys with other inexpensive metals.[8,25] Previous studies have revealed that Pd-based BMG catalysts, such as nano-structured $Pd_{40.5}Ni_{40.5}Si_{4.5}P_{14.5}$ BMG and $Pd_{40}Ni_{10}Cu_{30}P_{20}$ MG ribbons, have two unusual characteristics: (i) a self-stabilizing behavior under long-term static overpotentials and (ii) an improved activity as the number of cyclic voltammetry (CV) cycles increases due to selective dealloying.[9,26–28] These characteristics of Pd-based BMG catalysts are beneficial compared to the deteriorating activity found for polycrystalline Pd-alloys.[9,12,26,28–32] Most electrocatalytic research on Pt-based BMGs focuses on hydrogen oxidation reactions for fuel cells.[12,23,24,33] The puzzle of Pt-based BMGs for HER in hydrogen production is still unexplored.

This work explores the electrocatalytic performance of $Pt_{57.5}Cu_{14.7}Ni_{5.3}P_{22.5}$ BMG for hydrogen production upon HER. $Pt_{57.5}Cu_{14.7}Ni_{5.3}P_{22.5}$ BMG specimens with flat, micro-patterned, and nano-patterned surfaces were created through TPF. Superior electrocatalytic properties of nanorod-patterned Pt-BMG were observed in linear sweep voltammetry (LSV), Tafel plots, electrochemical impedance spectroscopy (EIS), and cyclic voltammograms. Furthermore, nanorod-patterned Pt-BMG shows a remarkable improvement as the number of LSV cycles increases in the stability tests. Scanning electron microscopic (SEM) images, X-ray photoelectron spectroscopy (XPS), and Auger electron spectroscopy (AES) reveal that a layer of $CuO/Cu_2O$ ($Cu_xO$) foam forms after 1000 cycles, covering the $Pt_{57.5}Cu_{14.7}Ni_{5.3}P_{22.5}$ nano-patterns. A three-step process is proposed to explain the formation of the $Cu_xO$ foam without using copper salt via DHBT electrodeposition from the Cu dissolution of Pt-BMG. Overall, our results reveal that nano-structured Pt-based BMG opens a new possibility of creating high-efficiency and long-term stability electrocatalysts for hydrogen generation.

## 2. Materials and Methods

### 2.1. Casting of $Pt_{57.5}Cu_{14.7}Ni_{5.3}P_{22.5}$ BMG

$Pt_{57.5}Cu_{14.7}Ni_{5.3}P_{22.5}$ BMG was produced via induction-melting and tilt-casting. High-purity elements (> 99.95%) were used to alloy a feedstock of $Pt_{57.5}Cu_{14.7}Ni_{5.3}P_{22.5}$ (Pt-BMG) in an induction furnace, where the phosphorus losses due to evaporation were minimal. The feedstock was cast into a Pt-BMG rod with 6 mm diameter and 100 mm length by tilt-casting into a copper mold. No fluxing chemicals were added, and all melting and casting processes were carried out in a high-purity argon protective environment. The as-cast rod was cut into disks with 0.8 mm thickness. The disks from the bottom, middle, and top sections of the as-cast rod were characterized by X-ray diffraction (XRD) using a Bruker D2 Phaser diffractometer with Co Kα radiation ($\lambda$ = 0.17889 nm) to prove the amorphous structure of the whole as-cast rod. No crystallinity was detected in any of the Pt-BMG disks. The disks were ground and mirror-polished to a thickness of roughly 0.7 mm to eliminate potential surface oxidation and to guarantee parallel surfaces for TPF processing.

### 2.2. Thermoplastic Forming of $Pt_{57.5}Cu_{14.7}Ni_{5.3}P_{22.5}$ BMG

The TPF apparatus utilized in this study was inspired by previous work by Schroers et al.[34,35] The setup is based on a compression machine (Zwick Z100, ZwickRoell GmbH & Co. KG) with ad-hoc components, featuring upper and lower anvils fitted with heating cartridges, a PID controller for process control, and water-cooling circulation to protect the load cells from the heat. A polished $Pt_{57.5}Cu_{14.7}Ni_{5.3}P_{22.5}$ BMG disk with a diameter of 6 mm and a thickness of around 0.7 mm was placed on a template. The Pt-BMG disk was then heated to the SCLR (253 ± 2°C) and imprinted from the template. Table 1 shows the TPF processing parameters. Pt-BMG samples with different surface topographies of flat, micro-rods, and nano-rods were fabricated using different templates. Flat surfaces served as references and were imprinted using a mirror-polished (roughness less than 1 μm) stainless steel sheet foil as template. Micro-rods were imprinted using macro-porous silicon templates (Si mold) with 2.5 μm pore diameter and 4.2 μm interpore distance (SmartMembranes GmbH). Nano-rods were imprinted from nano-porous alumina templates (AAO mold) with 90 nm pore diameter and 125 nm interpore distance (SmartMembranes GmbH). Silicon templates were dissolved in KOH solution (1.5 M), while alumina templates were dissolved in phosphoric acid solution (5% w/w).

**Table 1:** TPF processing parameters.

| Nomenclature | Topography | Max. Force [kN] | Loading Rate [kN min$^{-1}$] | Loading Time [Minute] | Holding Time [Minute] |
|---|---|---|---|---|---|
| Flat | Flat | 5.6 ± 0.2 | 5.6 | 1 | 10 |
| Micropattern (Micro) | Micro-rods | 1.4 ± 0.2 | 0.7 | 2 | 4 |
| Nanopattern (Nano) | Nano-rods | 5.6 ± 0.2 | 5.6 | 1 | 6 |

## 2.3. Electrochemical Measurements

The electrochemical measurements were conducted in a three-electrode glass cell at room temperature. A RE-1BP type Ag/AgCl reference electrode with a ceramic junction filled with 3 M KCl electrolyte (+0.195 V vs. reference hydrogen electrode (RHE)) and a 0.5 mm diameter 23 cm long Pt wire ring were used as reference and counter electrodes, respectively. The samples for the electrochemical measurements were subsequently sonicated in acetone, isopropyl alcohol, and water for 5 minutes each, then dried by hot air. For the micro- and nano-patterned samples, in order to eliminate the signal of the back side, the back side of the samples was painted with varnish. The electrochemical measurements were carried out with a Gamry Interface 1010E Potentiostat/Galvanostat/ZRA. Samples were submerged into 0.5 M $H_2SO_4$ solution. Once the open circuit potential (OCP) became steady (i.e., maximum of 5 mV change with 1 h), and electrochemical impedance spectroscopy (EIS) was performed at OCP (─0.356 V for flat, ─0.428 V for micro-patterned, and ─0.306 V for nano-patterned samples, respectively) at an AC amplitude of 0.01 V recorded from 100000 to 0.1 Hz. Electrochemically active surface area (ECSA) of flat, micro, and nanofeatured samples were measured by cyclic voltammetry (CV) before and after HER measurements. Subsequently, linear sweep voltammetry (LSV) was conducted with the same electrodes at a scan rate of 0.005 V s$^{-1}$ with potentials starting from ─1.2 V to 0.3 V with a shift in the onset potential of 0.4 V at every cycle until ─0.2 V. Among these measurements, LSV from ─1.0 V to 0.3 V was displayed, and Tafel slope values were calculated from these curves, which were drawn with respect to the ECSA calculations. The change in ESCA after 1 LSV cycle is extremely small (within error limits of %5) and hence can be ignored. After LSV was finished at the 1$^{st}$ and 1000$^{th}$ cycle, EIS was performed again under the same conditions. Suitable electrochemical circuit models (ECMs) were built and simulated in ZSimpWin V 3.10 (AC Impedance data analysis software). In order to measure the electrochemically active surface area (ECSA), the cyclic voltammetry (CV) scan was done with the TPF-prepared samples between ─0.1 V and 1.2 V (negative to positive) at a scan rate of 0.02 V s$^{-1}$. Stability measurements were performed on a fresh nano-patterned sample between

─0.8 V and 0.2 V at 0.02 V s$^{-1}$ for 1000 LSV cycles. The last measurement was done with 0.005 V s$^{-1}$ to measure the Tafel slope.

### 2.4. Surface Characterization

Overview images of the sample disks were recorded with confocal laser scanning microscopy (CLSM-Olympus LEXT OLS4100). The surface topographies of Pt-BMG specimens were imaged with 30° tilt via scanning electron microscopy (SEM- Zeiss LEO type 1525) using an In-Lens detector. The cross-section for SEM imaging was prepared using an Ion Milling System (E-3500, Hitachi) at an acceleration voltage of 6 kV and a discharge voltage of 4 kV for 8 hours. The ion-polished cross-sections were investigated by SEM (SEM Magna, Tescan) and energy dispersive X-ray spectroscopy (EDX; XFlash 6-60, Bruker corporation). X-ray photoelectron spectroscopy (XPS) analysis was carried out to investigate the chemical composition of the passive films formed on nano-patterned samples after 1 and 1000 LSV cycles (a flat sample was used as reference). To capture XPS spectra, a PHI 5600 spectrometer (Physical Electronics) with an Al Kα monochromatic X-ray source (200 W) was used. For high-resolution scans, a pass energy of 29 eV was applied. A neutralizer was engaged to avoid charging effects. The binding energy is referenced to C1s 284.8 eV. The collected XPS data was fitted using the Unifit software. Auger electron spectroscopy (AES) analysis was implemented on the nano-patterned sample after 1000 LSV cycles for chemical analysis of the surface. The spectrometer (JEOL JAMP 9500F, Tokyo, Japan) is equipped with a hemispherical analyzer and operated with 10 kV and 10 nA electron beam conditions. The samples were sputter-cleaned for 30 seconds prior to the AES measurements to eliminate possible contamination. The argon ions were accelerated with 2.1 keV, leading to an estimated sputter rate of 6 nm/min (calibrated to 100 nm SiO$_2$ from Si).

### 3. Results

### 3.1. Surface Topography of the Pt-BMG Electrocatalysts

The Pt$_{57.5}$Cu$_{14.7}$Ni$_{5.3}$P$_{22.5}$ BMG electrocatalysts with flat, micro-patterned, and nano-patterned surfaces were fabricated via the TPF technique. The resulting samples are shown in Figure 1. The flat samples (Figures 1(a-c)) have a diameter of around 11 ± 0.1 mm and a thickness of 0.16 ± 0.1 mm and a smooth surface topography. The micro-patterned samples (Figures 1(d-f)) have a diameter of about 7.6 ± 0.1 mm and a thickness of 0.20 ± 0.1 mm, with a mico-rod array surface feature. The micro-rods are 2.5 μm in diameter with a 4.2 μm inter-rod distance, and a 1.7 μm gap between each other. The nano-patterned samples (Figures 1(g-i)) have a diameter of around 10.5 ± 0.1 mm and a thickness of 0.17 ± 0.1 mm, and exhibit a nano-rod array surface features. The nano-rods have a diameter of 90 nm, a 125 nm inter-rod distance, and a 35 nm gap between each other.

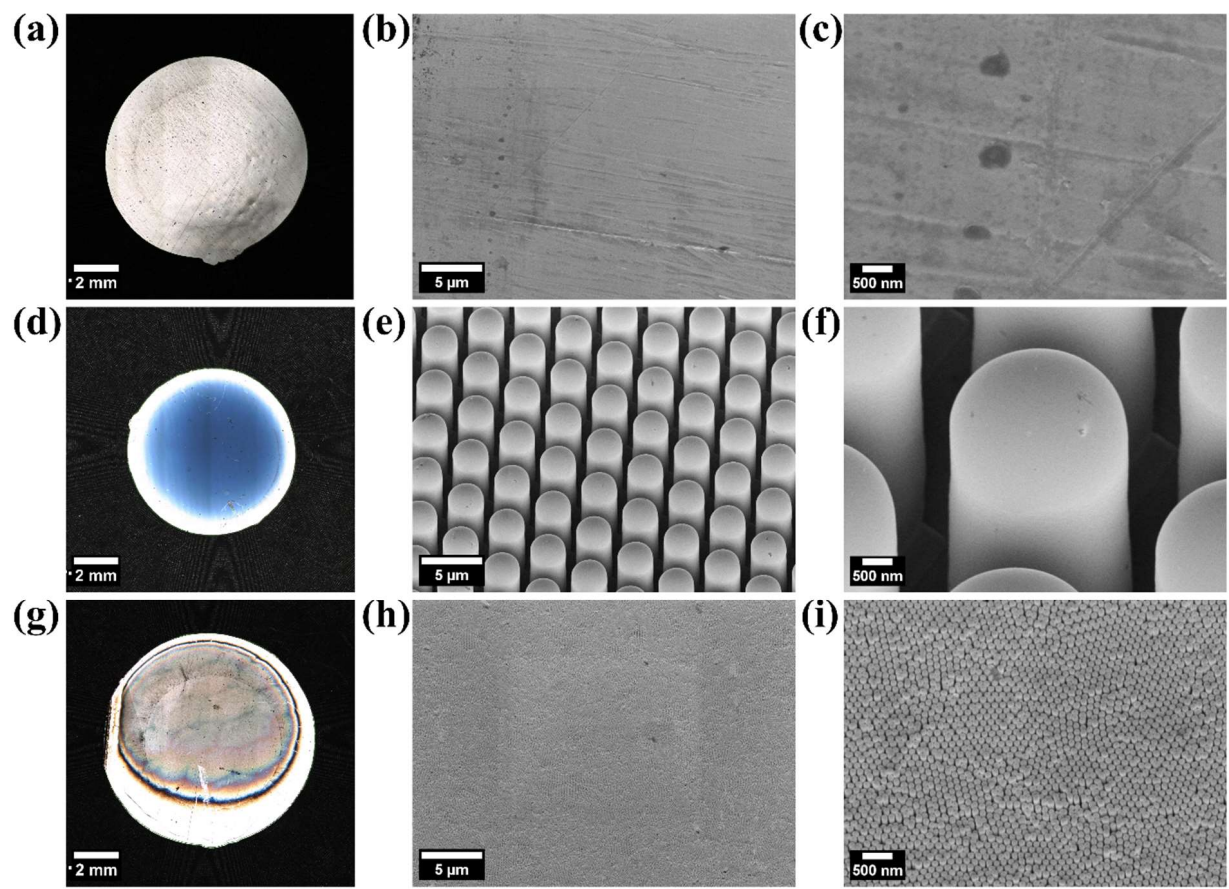

**Figure 1:** Thermoplastic formed $Pt_{57.5}Cu_{14.7}Ni_{5.3}P_{22.5}$ BMG with (a-c) a flat surface, (d-f) a micro-rod surface, and (g-i) a nano-rod surface. (a)(d)(g) show overview images of the entire BMG disk from CLSM, while (b)(e)(h) and (c)(f)(i) exhibit SEM images at 10k and 50k magnification, respectively.

### 3.2. Electrochemical Hydrogen Evolution Kinetics

In order to reveal the electrochemical hydrogen evolution kinetics of flat, micro-patterned, and nano-patterned samples, LSV cycles were performed (Figure 2(a)). The ECSA values of flat, micro-patterned, and nano-patterned samples are 1.936 mm², 2.886 mm², and 2.598 mm², respectively. Based on the ECSA calculations, the overpotential $E$ (V vs. RHE) required to deliver a current density of $J = -10$ mA cm$^{-2}$ for the flat, micro-patterned, and nano-patterned samples is —0.48 V, —0.37 V, and —0.30 V, respectively. All samples reached chemical equilibrium close to 0 mA cm$^{-2}$ at ~ —0.1 V. Compared to the flat samples (145 ± 2 mV dec$^{-1}$), the Tafel slopes determined from the relatively straight region of the cathodic region evidence a decrease for the micro-patterned samples (86 ± 2 mV dec$^{-1}$), as shown in Figure 2(b). The Tafel slope further decreases to 67 ± 1 mV dec$^{-1}$ for the nano-patterned samples. This reveals that a smaller overpotential range is needed to change the current density by one order of magnitude (decade), confirming an acceleration of the hydrogen evolution reaction.[36,37]

The stability of the electrodes can be measured using electrochemical cycling performance tests. The nano-patterned samples were selected due to their lowest Tafel slope and smallest absolute overpotential for HER. Figure 2(d) depicts the evolution of the LSV curves of the nano-patterned samples under 1000 cycles. Up to 50 cycles, the absolute overpotential |E| increases slightly. However, after this point, the absolute overpotential |E| decreases significantly for 1000 cycles, corresponding to the overpotential required to achieve a current density of $J = -10$ mA cm$^{-2}$ shifts from —0.26 V at the 1st cycle to —0.15 V at the 1000th cycle. The ECSA values increased from 1.984 mm$^2$ to 3.247 mm$^2$ after 1000 LSV cycles. The decrease in the absolute overpotential also affects the Tafel slopes (Figure 2(e)), which vary from 67 ± 1 mV dec$^{-1}$ to 42 ± 1 mV dec$^{-1}$. These remarkable small values are comparable to the benchmark materials in HER electrocatalysis (Table S1).[38–40]

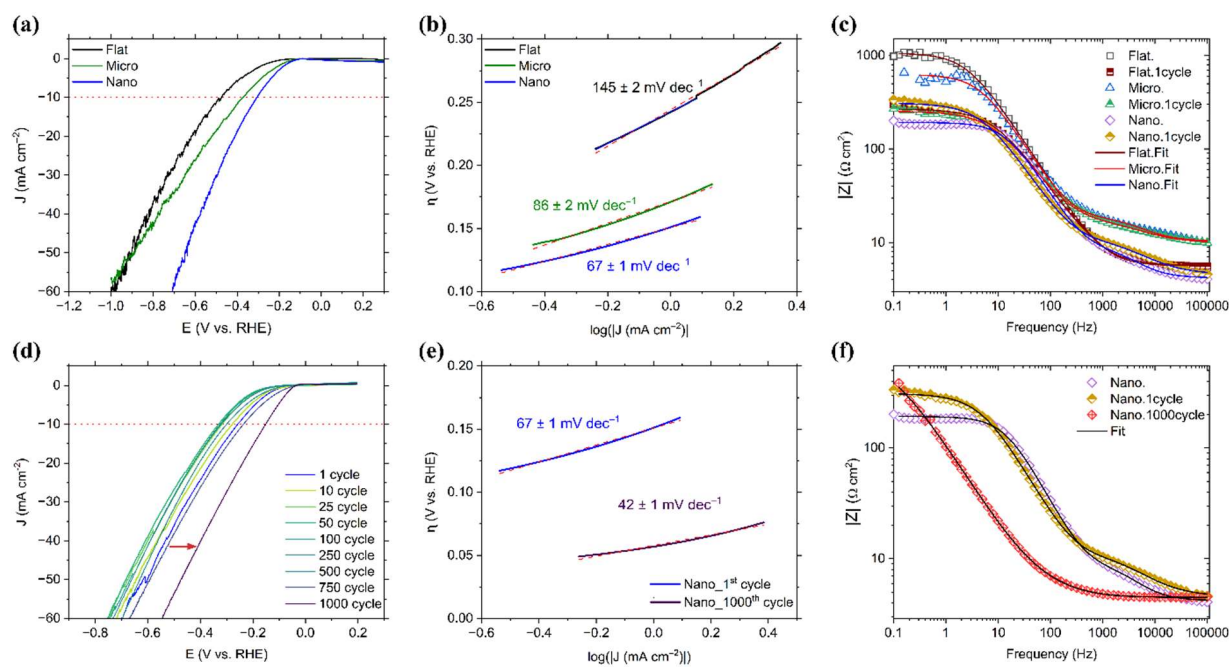

**Figure 2:** Results of electrochemical studies of Pt$_{57.5}$Cu$_{14.7}$Ni$_{5.3}$P$_{22.5}$ BMG electrocatalysts in 0.5 M H$_2$SO$_4$ electrolyte: (a) LSV of the flat, micro-patterned and nano-patterned samples, and (b) their corresponding Tafel curves. (c) The Bode magnitude plot of the samples before and after 1 cycle LSV. (d) Electrochemical cycling performance tests for 1000 cycles. (e) The Tafel curves of the nano-patterned sample in the 1st and 1000th cycle. (f) The Bode magnitude plot of nano-patterned samples between fresh (Nano.), one cycle (Nano.1cycle) and 1000th cycle (Nano.1000cycle) states.

The nano-patterned Pt-BMG shows a self-improvement behavior in electrocatalytic performance after 1000 LSV cycles. Additional stability tests for 200 LSV cycles (with samples positioned horizontally) were conducted to compare the self-improvement behavior of flat, micro-patterned, and nano-patterned Pt-BMGs. Figure S1 shows that after 200 LSV cycles, only the nano-patterned sample has the absolute overpotential |E| decreased by 8% for $J = -10$ mA cm$^{-2}$. In contrast, the absolute overpotential |E| of the flat and micro-patterned samples increases by 41% and 19%, respectively. Therefore, nano-patterned Pt-BMGs outperform their flat and micro-patterned counterparts regarding lower absolute overpotential, smaller Tafel slope value, and long-term stability for HER applications.

The Bode magnitude plot in Figure 2(c) shows distinct differences before and after the first LSV cycle. For flat and micro-patterned samples, decreases in the magnitude |Z| at the mid-to-low frequency range (<1000Hz) are observed. Interestingly, for the nano-patterned samples, the magnitude |Z| at the near-DC range (<10 Hz) increases after the first LSV cycle. However, in Figure 2(f), the magnitude |Z| of the 1000th cycle (Nano.1cycle) is lower than fresh (Nano.) and one cycle (Nano.1cycle) states, which is the case for almost the entire overall frequency range, yet approaching the one cycle state at the lowest frequency.

### 3.3. Analysis of an Equivalent Circuit Model (ECM) from EIS Measurements

The EIS data from Figure 2(c, f) were further fitted by a R(QR) equivalent circuit model (ECM) for the flat samples and by a R(Q(R(QR))) ECM for the micro- and nano-patterned samples (Table 2). The R(QR) model for the flat samples includes a solution resistance $R_s$, a charge transfer resistance $R_{ct}$, and a constant phase element (CPE) $Q_{dl}$ for the double-layer capacitance. Besides $R_s$, $R_{ct}$, and $Q_{dl}$, the R(Q(R(QR))) model for the micro- and nano-patterned samples adds a patterned layer CPE $Q_{pl}$ and a patterned layer resistance $R_{pl}$ to enhance the fit quality as well as to analyze the influence of surface modifications due to patterning. It is worth noting that CPE is used to account for the non-ideal behavior of the charged interface.[41,42] The impedance of CPE is defined as $Z_{CPE} = 1/Y_0(j\omega)^n$, where $Y_0$ is CPE parameter and $n$ is CPE exponent.[43] Hence, $Q_{dl}$-$Y_0$ and $Q_{dl}$-$n$ are the CPE parameter and the CPE exponent of the double-layer, respectively. $Q_{pl}$-$Y_0$ and $Q_{pl}$-$n$ are the CPE parameter and the CPE exponent of the patterned layer, respectively.

**Table 2:** ECM of the samples before and after LSV cycles. $R_s$: solution resistance, $Q_{dl}$-$Y_0$: constant phase element (CPE) parameter of the double-layer, $Q_{dl}$-$n$: CPE exponent of the double-layer, $R_{ct}$: charge transfer resistance, $Q_{pl}$-$Y_0$: CPE parameter of the patterned layer, $Q_{pl}$-$n$: CPE exponent of the patterned layer, $R_{pl}$: resistance of the patterned layer, $\chi^2$: chi-squared.

| Sample | Circuit Model | $R_s$ ($\Omega$ cm$^2$) | $Q_{dl}$-$Y_0$ (S s$^n$ cm$^{-2}$) | $Q_{dl}$-$n$ (-) | $R_{ct}$ ($\Omega$ cm$^2$) | $Q_{pl}$-$Y_0$ (S s$^n$ cm$^{-2}$) | $Q_{pl}$-$n$ (-) | $R_{pl}$ ($\Omega$ cm$^2$) | $\chi^2$ |
|---|---|---|---|---|---|---|---|---|---|
| Flat. | 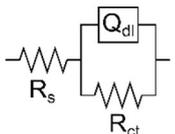 | 5.6 | $2.2 \times 10^{-5}$ | 0.87 | 1064 | - | - | - | $2.1 \times 10^{-3}$ |
| Flat. 1cycle | | 5.7 | $3.8 \times 10^{-5}$ | 0.81 | 263 | - | - | - | $1.3 \times 10^{-3}$ |
| Micro. | 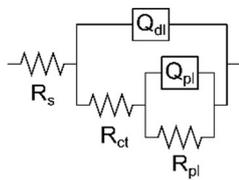 | 10.3 | $7.0 \times 10^{-6}$ | 0.91 | 808 | $2.9 \times 10^{-6}$ | 0.83 | 10.4 | $9.2 \times 10^{-3}$ |
| Micro. 1cycle | | 9.9 | $1.4 \times 10^{-5}$ | 0.82 | 236 | $6.8 \times 10^{-6}$ | 0.76 | 9.0 | $1.1 \times 10^{-3}$ |
| Nano. | | 4.2 | $8.5 \times 10^{-6}$ | 0.92 | 182 | $5.0 \times 10^{-6}$ | 0.88 | 6.6 | $1.8 \times 10^{-3}$ |
| Nano. 1cycle | | 4.5 | $1.5 \times 10^{-5}$ | 0.89 | 299 | $1.4 \times 10^{-5}$ | 0.73 | 8.1 | $9.5 \times 10^{-4}$ |
| Nano. 1000cycle | | 4.8 | $1.0 \times 10^{-3}$ | 0.71 | 649 | $1.3 \times 10^{-3}$ | 0.79 | 7.5 | $1.4 \times 10^{-3}$ |

As illustrated in Figure 2(c, f), the fitting curves exhibit a high degree of correlation with the Bode magnitude plots derived from the EIS data within the measured frequency range. The solution resistance $R_s$ deviates very slightly, and the values are negligible, confirming that the system is mainly influenced by the electrochemical reactions. For the double-layer capacitance $Q_{dl}$, there is an increase in the CPE parameter $Q_{dl}$-$Y_0$ for all the samples, whereas the CPE exponent $Q_{dl}$-$n$ decreases slightly after HER. The significant decreases in the charge transfer resistance $R_{ct}$ for the flat and micro-patterned samples dominate their decreases in the magnitude of the impedance in Figure 2(c).

The patterned layer CPE $Q_{pl}$ in the circuit reflects the response of the micro- and nano-patterns. The CPE parameter $Q_{pl}$-$Y_0$ of the micro-patterns increases significantly with a dramatic decrease in the relevant CPE exponent $Q_{pl}$-$n$. $R_{pl}$ varies slightly, but the values are generally quite low, revealing that electron transfer happens extremely fast on such patterned surfaces.

Regarding the nano-patterned samples, especially after 1000 cycles, $Q_{dl}$-$Y_0$ and $Q_{pl}$-$Y_0$ increase by two orders of magnitude after 1000 LSV cycles, which could be attributed to the ion accumulation. The notable decrease in $Q_{dl}$-$n$ implies a substantial deviation from ideal capacitive behavior. $R_{ct}$ increases by more than twofold, suggesting increased difficulty for ions to cross the electrode-electrolyte interface. These observations provide support for the subsequent conclusion that a porous layer has formed on top of the Pt-BMG nanorods after 1000 LSV cycles.

## 3.4. Surface Characterization After 1000 Cycles

As shown in Figure 3, the surfaces of the nano-patterned samples were characterized after 1 and 1000 LSV cycles by multiple complementary methods to find the reason for the improved electrocatalytic performance. X-ray photoelectron spectroscopy (XPS) allows for the analysis of the elemental chemical states on the material surface. The Pt4f spectrums are revealed in Figure 3(a), where sputter-cleaned and untreated Pt foils are provided as references for the Pt-metal and Pt-oxide, respectively. The Pt-metal has a characteristic doublet at ∼71 and ∼74 eV, while Pt-oxide has a second doublet at ∼72 and ∼75 eV.[23,44] Hence, the Pt in the Pt-BMG after 1 LSV cycle is in a metallic state, similar to the sputter-cleaned Pt foil, and the slightly different FWHMs are due to the more heterogeneous environment in the BMG compared to the pure metal. Interestingly, no Pt signal is detected on the BMG surface after 1000 LSV cycles.

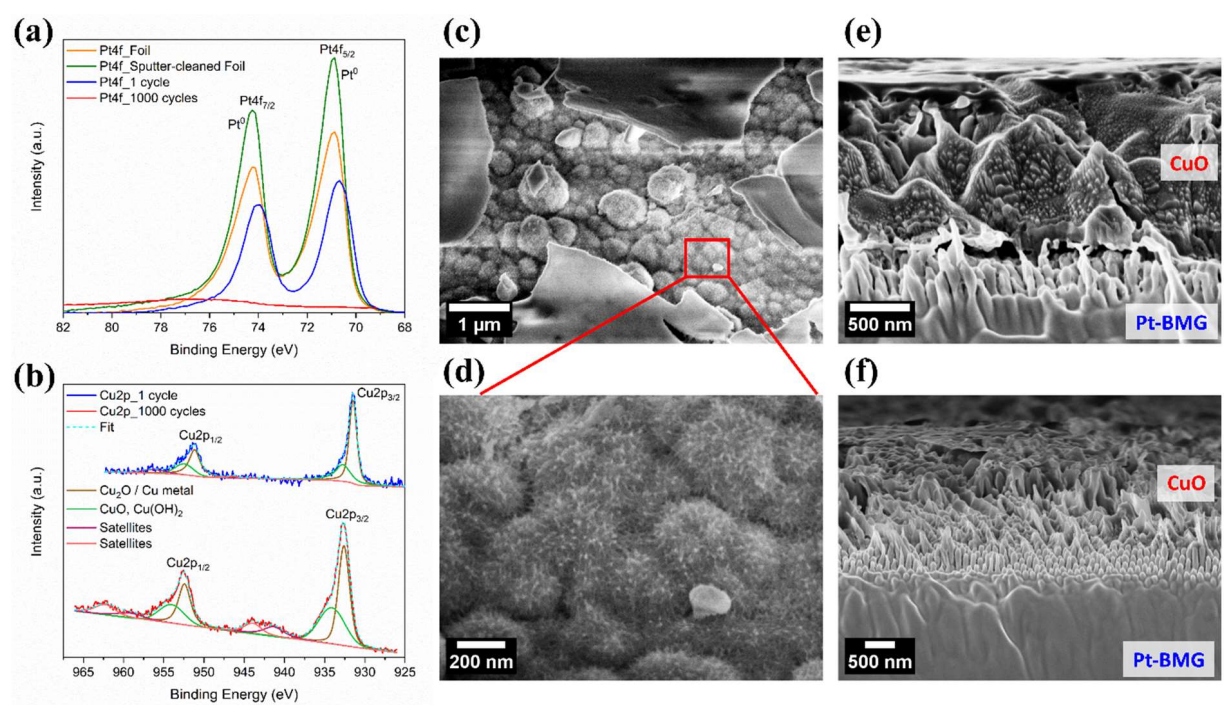

**Figure 3:** Surface characterization of nano-patterned $Pt_{57.5}Cu_{14.7}Ni_{5.3}P_{22.5}$ BMG after 1000 cycles: (a)(b) XPS spectra of (a) Pt4f and (b) Cu2p. (c)(d) SEM images of the top view where (d) is the zoomed-in view of the dandelion-like CuO in (c). (e)(f) SEM images of the cross-section view showing dandelion-like CuO and smooth islands like CuO in (d) and CuO nanosheets in (e).

Regarding the Cu2p spectra, it has been reported that Cu-metal and $Cu^+$ have signals at ~932 and ~952 eV, whereas $Cu^{2+}$ has signals at ~934 and ~954 eV.[17,45–47] As shown in Figure 3(b), the BMG surfaces after both 1 LSV cycle and 1000 LSV cycles show prominent signals associated with Cu-metal/$Cu_2O$ (at ~932 and ~952 eV) and CuO (at ~934 and ~954 eV) while more substantial satellite peaks indicating a mixture of CuO and $Cu(OH)_2$ are found for the sample after 1000 cycles. It is worth noting that the Cu2p signals for $Cu_2O$ or Cu-metal are challenging to distinguish in this case. Generally, the Cu content is higher after 1000 cycles than after 1 cycle, and no Pt is detected after 1000 cycles. Even after 1 LSV cycle, Ni and P signals were not detected from XPS on the surface of the flat and the nano-patterned Pt-BMGs. AES was used to clarify the composition of the surface of the nano-patterned Pt-BMG after 1000 LSV cycles. Figure S2(a) provides a representative SEM image with markers of the local AES probing areas, and Figure S2(b) displays the spectra of Pt, Cu, C, and O. The forming surface foam on the nano-patterned Pt-BMG after 1000 LSV cycles contains CuO rather than Pt-BMG, supporting the XPS findings.

The surface topography of nano-patterned Pt-BMG changes dramatically after 1000 LSV cycles, as displayed in Figure 3(c). The Pt-BMG nanorods shown in Figure 1(i) are covered by $CuO/Cu_2O$ with smooth island-like and dandelion-like topographies. The dandelion-like topography is composed of $CuO/Cu_2O$ with fur-like structures, leading to a porous structure, as revealed in Figure 3(d). Due to the high corrosion resistance of Pt in the electrolyte used, it is unlikely that the Pt-BMG nanorods dissolved in the electrolyte during the stability test. Hence, the ion-polished cross-section of the sample was inspected by SEM (illustrated in Figures 3(d-e)), revealing that an about 1 μm thick layer of $CuO/Cu_2O$ foam forms on top of the Pt-BMG nanorods. Several agglomerated CuO particles with only a few microns in size form above the deposited layer (Figure S3). The following topographies were found for the deposited $CuO/Cu_2O$ foam: (1) Smooth island-like CuO in Figure 3(c), (2) dandelion-like CuO in Figure 3(d), and (3) CuO nanosheets in Figure 3(e).

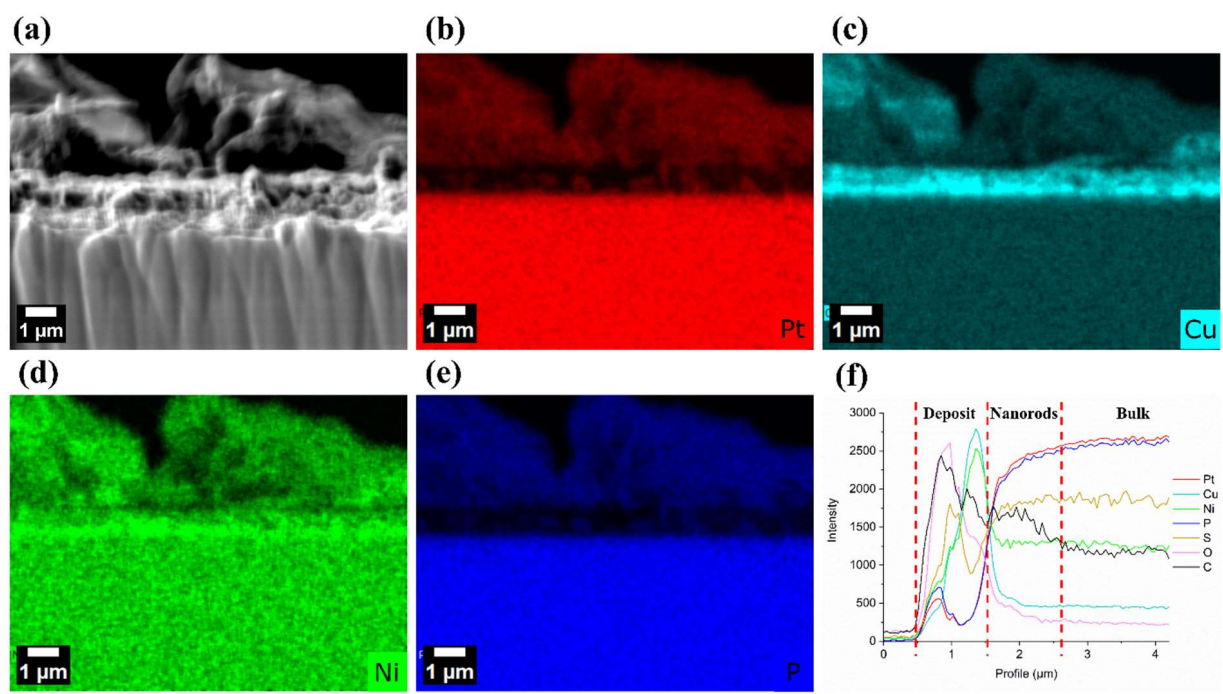

**Figure 4:** EDX elemental mapping of the cross-section of the nano-patterned Pt-BMG after 1000 LSV cycles. (a) The representative SEM image. (b-e) Elemental mapping of (b) Pt, (c) Cu, (d) Ni, and (e) P. (f) EDX line scan profile of (a).

The cross-section and the related EDX elemental distributions of the nano-patterned Pt-BMG after 1000 LSV cycles are shown in Figure 4. The EDX mapping confirms that the newly deposited layer mainly comprises Cu, as illustrated in Figure 4(c). It is worth noting that the elemental distribution also shows high Ni contents in the deposited layer, which is an artifact of the overlapping spectrum with Cu (Figure 4(d)). In Figure 4(f), the line scan profile of the cross-section reveals that the newly deposited layer has a thickness of around 1 μm. The upper deposit layer is mainly composed of C and O, which can be attributed to carbon contamination from hydrocarbon adsorption of the atmosphere. In contrast, the lower deposit layer mainly consists of Cu, which belongs to the CuO/Cu$_2$O layer identified by XPS and AES analysis.

## 4. Discussion

The efficiency of HER is higher in an acidic electrolyte because of an abundance of protons accessible for instantaneous charge and discharge in the acidic medium compared to an alkaline medium.[38] However, weak stability and insufficient corrosion resistance of electrocatalysts are usually the main challenges for commercial-scale applications with long-term operation times.[38,39,48] Therefore, an ideal electrocatalyst for water splitting should have a low overpotential, a low Tafel slope value, a large exchange current density, and long-term stability.[38,48] The nano-patterned Pt-BMGs have lower overpotential and smaller Tafel slopes than their flat and micro-patterned counterparts. Moreover, in the stability test, the nano-patterned Pt-BMGs demonstrate self-improvement characteristics in electrocatalytic performance after 1000 LSV cycles. Combining the findings of the EIS, XPS, AES, SEM, and EDX analyses indicates that a $Cu_xO$ layer forms on the top of Pt-BMG nanorods after 1000 LSV cycles.

Firstly, the EIS data and ECM fitting of the nano-patterned Pt-BMG show that after 1000 LSV cycles compared to 1 LSV cycle, (1) the CPE parameters of the double-layer ($Q_{dl}$-$Y_0$) and the patterned layer ($Q_{pl}$-$Y_0$) increase by two orders of magnitude, and (2) the charge transfer resistance ($R_{ct}$) increases more than twice, implying the deposition of an porous layer on the surface (Table 2). XPS and AES analyses further show that this extra surface layer mainly consists of $Cu_xO$ from HER. Since the $Cu_xO$ layer is deposited on top of the Pt-BMG and covers the entire surface, surface-sensitive methods like XPS can not detect the Pt signal from the nano-patterned Pt-BMG after 1000 LSV cycles. Finally, the SEM images from the top view and the cross-section reveal that a $Cu_xO$ layer was deposited on top of the Pt-BMG nanorods. The formed $Cu_xO$ layer has a thickness of about 1 μm, and three types of topographies are observed: (1) Smooth islands, (2) dandelion-like structure, and (3) nanosheets. The EDX elemental distribution further confirms carbon contamination (from hydrocarbon adsorption of the atmosphere) of the upper deposition layer and a Cu-rich layer on the lower deposition side. The following sections discuss how the $Cu_xO$ layer was deposited and how it improves the HER.

## 4.1. How Is $Cu_xO$ Deposited on the Nano-patterned Pt-BMG During LSV Scans?

A three-step mechanism is proposed to explain the formation of a porous $Cu_xO$ layer on the nano-patterned Pt-BMG during LSV cycles (Figure 5): (i) dissolution of Cu from Pt-BMG in the low overpotential section, (ii) re-deposition of Cu foam via dynamic hydrogen bubbles templating (DHBT) electrochemical deposition in the high overpotential section, (iii) oxidation of the deposited Cu to $Cu_xO$ during the low overpotential section of the LSV cycle.

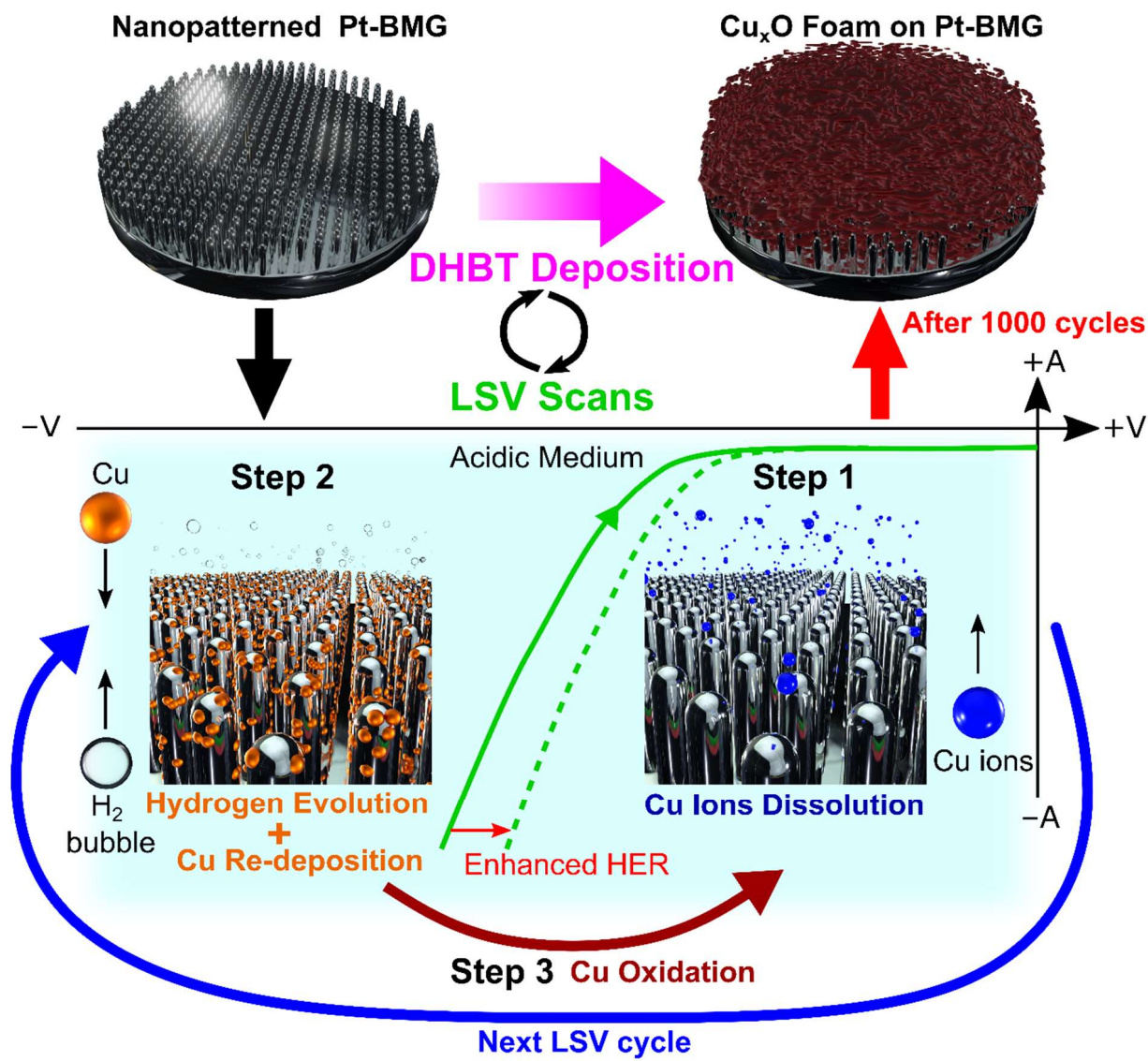

**Figure 5:** Proposed mechanism of Cu dissolution and CuO/Cu$_2$O foam formation on the nano-patterned Pt-BMG.

In the first step, Cu electro-dissolution occurs from the nano-patterned Pt$_{57.5}$Cu$_{14.7}$Ni$_{5.3}$P$_{22.5}$ BMG under cathodic polarization conditions during the low overpotential part of LSV scans, providing a source of Cu ions at the surface–electrolyte contact.[47] According to the conventional Pourbaix diagram of Cu, it is unlikely that Cu will dissolve under our experimental conditions (potential: —0.8 to 0.2 V vs. RHE; electrolyte: 0.5 M H$_2$SO$_4$). However, conventional Pourbaix diagrams are restricted to the bulk form of pure metals and have limitations in determining dissolution rates.[49] For instance, Taylor et al. illustrated a drastically negative shift of the dissolution balance potential in the Pourbaix diagram for Cu nanomaterials because a large number of surface atoms with low coordination in nanomaterials have strong adsorption with water, oxygen, hydrogen, and hydroxyl ions, which could decrease the ionization potentials of the surface Cu atoms, as shown in Figure S4(a).[50,51] Therefore, the surface morphology of the nano-patterned samples with 90 nm diameter nano-rods has a higher tendency to dissolve Cu from Pt-BMG into the surrounding electrolyte than the flat and micro-patterned (2.5 μm diameter micro-rods) samples. Moreover, Kreizer et al. reported that in acidic solutions with trace oxygen, Cu dissolution occurs under even milder cathodic polarization conditions (—0.40 to —0.70 V vs. the standard hydrogen electrode (SHE)).[47,52,53] Because the evolving hydrogen bubbles stir the electrode/interface layer and make the electrode more accessible to oxygen, the Cu dissolution rate is increased by the formation/dissolution of adsorption-complex $Cu_xO_{x/2}$.[47,52,53] Their studies also showed that oxygen exposure is critical for Cu dissolution because of the required intermediate products ($Cu_xO_{x/2}$), and Cu dissolution could not occur in deaerated media.[53] The synergy effect incorporating (a) the negative shift of the dissolution balance potential from the nano-pattern and (b) the oxygen exposure due to the perturbed electrode/electrolyte layer from the evolved hydrogen bubbles, facilitates the nano-patterned Pt-BMG to leach $Cu^{2+}_{(aq)}$ ions, providing the source for re-deposition in the next step.

In the re-deposition step, during the high overpotential section of LSV scans, the dissolved $Cu^{2+}_{(aq)}$ ions are reduced to the metallic form on the electrode ($Cu^{2+}_{(aq)} + 2e^- \rightarrow Cu_{(s)}$) while the evolving hydrogen bubbles act as a depositing template. The concept of applying the evolving hydrogen bubbles during HER as a dynamic template to obtain 3D metal foams is known as the dynamic hydrogen bubble template (DHBT) method.[16,17] When the HER occurs at a cathode with a high overpotential, metal electrodeposition and hydrogen evolution operate simultaneously; thus, hydrogen bubbles interfere with the contact between metal ions and the cathode, and metal ions can only be electrodeposited between bubbles, leading to porous metal structures.[16,17] It has been reported that the surface topography of electrocatalysts influences the evolution of hydrogen bubbles via its hydrophilicity to water molecules and its aerophobicity to gas bubbles, respectively.[25,54,55] It is observed during LSV cycles that the nano-patterned Pt-BMG evolves smaller and faster hydrogen bubbles than the flat Pt-BMG (watch supplementary videos). Therefore, the surface pattern of Pt-BMG could be utilized to control the bubble formation rate and the bubble size during HER, thereby allowing further tuning of the pore size and density of the deposited Cu upon DHBT electrodeposition.

Finally, there are two possible reasons why the deposited Cu is oxidized to $Cu_xO$ ($Cu_2O$/CuO). Firstly, during the vigorous HER, proton consumption at the electrode/electrolyte interface is substantial, leading to an increase in local pH.[47,56] Since copper oxides are stable in alkaline and neutral solutions, such an electro-dissolution mechanism would encourage the formation of copper oxides on the deposit.[47] The other reason can be illustrated by a Pourbaix diagram of copper with the dashed line ''a'', as shown in Figure S4(b).[57] At potentials above this dashed line ''a'' in hydrogen evolution, Cu oxidation could occur ($2Cu + H_2O \rightarrow Cu_2O + 2H^+ + 2e^-$).[57] With this Pourbaix diagram, the re-deposited Cu foam can further oxidize when the LSV scan proceeds to the low overpotential section in the stability test.

This three-step process provides a strategy to achieve a porous $Cu_xO$ layer without using copper salt in the plating bath, opening another potential application for BMGs. In principle, any nano-patternable BMG alloy that contains Cu and another element with a higher dissolution potential than Cu can form $Cu_xO$ foams using the proposed three-step process. Hence, future research could focus on investigating more cost-effective BMG systems with high fragility to reduce production costs. Beyond hydrogen evolution applications, it would be interesting to further investigate the porous $Cu_xO$ layer created in this work for other potential applications, such as gas sensing, supercapacitor, photocatalytic, and antibacterial applications.[46,58–60]

**4.2. How Does the Newly Deposited $Cu_xO$ foam Enhance the HER Performance?**

The nano-patterned Pt-BMG already shows superior electrocatalytic performance to the flat and micro-patterned Pt-BMGs during the initial LSV scans, as revealed in Figures 1 and S2. It is expected that the $Cu_xO$ layer has not formed yet during the initial LSV scans. Therefore, the superior electrocatalytic performance can be attributed to the surface topography of nano-rods. Nano-topographies have been reported to promote HER via hydrophilicity and aerophobicity.[25,61,62] Consequently, the nano-patterned surface renders the Pt-BMG more hydrophilic and aerophobic compared with flat and micro-patterned specimens. The hydrophilic and aerophobic surfaces reduce the adhesion between hydrogen bubbles and the material surface, allowing bubbles to escape from the surface with a smaller bubble size.[63] This enables rapid departure of the evolved hydrogen bubbles from the catalyst while preventing the bursting of large hydrogen bubbles (watch supplementary videos), which could deteriorate the structural integrity and stability of the catalyst.[63] Moreover, hydrophilic surfaces generally possess higher mass transfer efficiency, increasing the possibility of collisions between water molecules and catalytic active sites.[63] The zigzag nature of the LSV curves indicates hydrogen bubble formation, growth, and detachment. In Figure 2(a), the flat Pt-BMG displays more prominent zigzags in the LSV curve, indicating significant coalescence of hydrogen bubbles. On the contrary, the nano-patterned Pt-BMG has a smoother LSV curve than the flat Pt-BMG, implying the evolution of small hydrogen bubbles on the nano-patterned surface. Thus, the nano-patterned surface has improved hydrophilicity and aerophobicity, elevating the turnover frequency of catalytic active sites that contribute to the HER. Further studies regarding bubble formation, growth, and detachment are required.

As the LSV cycles progress during the stability test, the newly deposited $Cu_xO$ foam builds on the nano-pattern of the Pt-BMG, further improving the HER performance. The impressive HER performance is attributed to the porous structures deposited, as discovered by SEM. These structures can strongly improve the ionic diffusion and electron transfer kinetics by increasing the surface area to volume ratio. As a result, the overpotential required to achieve a current density of $J = -10$ mA cm$^{-2}$ decreases by 42%, and the Tafel slope value is reduced by 37%.

Several studies have reported that CuO- and $Cu_2O$-derived materials exhibit respectable electrocatalytic and photocatalytic properties for HER.[59,64–69] Other studies based on Pd-based MGs, amorphous TiCuRu alloys for HER, and Pt-BMGs for ethanol oxidation reaction (EOR) show self-stabilizing or self-improvement characteristics in catalytic performance via selective dealloying of Cu.[23,25,27,70,71] Present work demonstrates that the dissolution of Cu and re-deposition of $Cu_xO$ foam on nano-patterned Pt-BMGs can be a strategy to enhance the catalytic activity and long-term stability for HER, uncovering a new mechanism for improving the durability of catalysts. As shown in Figure 2(d), the electrocatalytic performance improved significantly from 750 to 1000 cycles. The deposited $Cu_xO$ foam exhibits significant self-improvement with increasing LSV cycles. This indicates a potential for hydrogen evolution reaction (HER) applications, necessitating further in-depth studies on the long-term performance of nano-patterned Cu-containing BMGs.

## 5. Conclusion

This study focuses on the HER activity of flat and patterned $Pt_{57.5}Cu_{14.7}Ni_{5.3}P_{22.5}$ BMGs in an acidic environment. The results confirm a minimum Tafel slope of $67 \pm 1$ mV dec$^{-1}$ for the nano-patterned BMG. After 1000 LSV cycles of the same sample, the Tafel slope decreases by 37%, with an overpotential reduction of 42%. The HER performance is attributed to a $Cu_xO$ foam which forms on the nano-patterned surface, as confirmed by SEM, XPS, and EDX. This foam increases the surface area-to-volume ratio and provides more active sites, thereby enhancing the ionic diffusion and electron transfer kinetics significantly

A three-step process involving (i) dissolution of Cu from Pt-BMG in the low overpotential section, (ii) re-deposition of Cu foam via dynamic hydrogen bubbles templating (DHBT) deposition in the high overpotential section, and (iii) oxidation of the deposited Cu to $Cu_xO$, is proposed to explain how the $Cu_xO$ foam forms on the nano-patterned Pt-BMG during the stability test. Metallic glass is utilized in DHBT electrodeposition with the help of nano-patterning to fabricate porous metal-oxide foam without the need for metal salt. The resulting nanostructured catalyst exhibits high HER performance. More possibilities combining metallic glasses, thermoplastic patterning, and dynamic bubble templating as a new strategy for fabricating porous metal-oxide foams and self-improving catalysts still remain to be explored.

## Supporting Information

The Supporting Information is available free of charge at XXX.

The Tafel slope and overpotential η at 10mA $cm^{-2}$ of some reported electrocatalysts for HER; additional stability tests comparing the overpotential η (mV vs. RHE) at J = 10 mA $cm^{-2}$ of flat, micro-patterned and nano-patterned samples up to 200 LSV cycles; AES analysis of the Nano.1000cycle sample; SEM images of agglomerated CuO particles on top of the deposited CuO layer; modified Pourbaix diagrams for Cu with green arrows marking the experimental condition (PDF)

## Acknowledgements

This work was supported by the European Union's Horizon 2020 research and innovation program under the Marie Skłodowska-Curie grant agreement No. 861046 (BIOREMIA-ETN). B.S. J.E. and F.S. acknowledge support from the Austrian Science Fund (FWF), Grant/Award Number: I3937-N36.## References

(1) Pareek, A.; Dom, R.; Gupta, J.; Chandran, J.; Adepu, V.; Borse, P. H. Insights into Renewable Hydrogen Energy: Recent Advances and Prospects. *Materials Science for Energy Technologies* **2020**, *3*, 319–327. https://doi.org/10.1016/j.mset.2019.12.002.
(2) Filippov, S. P.; Yaroslavtsev, A. B. Hydrogen Energy: Development Prospects and Materials. *Russ. Chem. Rev.* **2021**, *90* (6), 627–643. https://doi.org/10.1070/RCR5014.
(3) Abe, J. O.; Popoola, A. P. I.; Ajenifuja, E.; Popoola, O. M. Hydrogen Energy, Economy and Storage: Review and Recommendation. *International Journal of Hydrogen Energy* **2019**, *44* (29), 15072–15086. https://doi.org/10.1016/j.ijhydene.2019.04.068.
(4) Yue, M.; Lambert, H.; Pahon, E.; Roche, R.; Jemei, S.; Hissel, D. Hydrogen Energy Systems: A Critical Review of Technologies, Applications, Trends and Challenges. *Renewable and Sustainable Energy Reviews* **2021**, *146*, 111180. https://doi.org/10.1016/j.rser.2021.111180.

## Supplementary Information

**Table S1:** The Tafel slope and overpotential η at 10mA cm$^{-2}$ of some reported electrocatalysts for HER.

| Catalysts | Tafel slope (mV dec$^{-1}$) | η (mV vs. RHE) at J = 10 mA cm$^{-2}$ | Stability | Electrolyte | Ref. |
|---|---|---|---|---|---|
| Nanorods patterned $Pt_{57.5}Cu_{14.7}Ni_{5.3}P_{22.5}$ BMG | 42 | 150 | 1000 cycles | 0.5 M $H_2SO_4$ | This work |
| Nanowires patterned $Pd_{40}Ni_{10}Cu_{30}P_{20}$ MG | 90.7 | 113 | - | 0.5 M $H_2SO_4$ | 1 |
| Pristine $Pt_3Te_4$ | 49 | 46 | - | 0.5 M $H_2SO_4$ | 2 |
| 200 nm-thick Pt film | 44.6 | 64.7 | - | 0.5 M $H_2SO_4$ | 3 |
| Polycrystalline Pt electrode | 31 | 73 | - | 0.5 M $H_2SO_4$ | 4 |
| Pt/C | 30 | 84 | - | 0.5 M $H_2SO_4$ | 4 |
| Pt sheet | 30 | 59 | - | 0.5 M $H_2SO_4$ | 5 |
| Pure Pt | 38 | - | - | 0.5 M $H_2SO_4$ | 6 |
| $Ir_{25}Ni_{33}Ta_{42}$ MG film | 35 | 99 | 1000 cycles | 0.5 M $H_2SO_4$ | 7 |
| Oxide-derived nanorods $Pd_{40.5}Ni_{40.5}Si_{4.5}P_{14.5}$ MG | 42.6 | 36 | 40000 sec | 0.5 M $H_2SO_4$ | 8 |
| $Pd_{40.5}Ni_{40.5}Si_{4.5}P_{14.5}$ MG plate | 73.6 | 138 | - | 0.5 M $H_2SO_4$ | 8 |
| $Ni_{20}Fe_{20}Mo_{10}Co_{35}Cr_{15}$ high entropy alloy | 41 | 107 | 8 hr | 0.5 M $H_2SO_4$ | 9 |
| $Pd_{20}Pt_{20}Cu_{20}Ni_{20}P_{20}$ high entropy MG | 44.6 | 62 | 100 hr | 0.5 M $H_2SO_4$ | 10 |
| Ni$_{4.5}$Fe$_{4.5}$S$_8$ in bulk | 72 | 280 | 1000 cycles | 0.5 M H$_2$SO$_4$ | 11 |
| PtCu/WO$_3$@CF | 45.9 | 41 | 2000 cycles | 0.5 M H$_2$SO$_4$ | 12 |
| CuO-modified stainless steel | 52 | 154 | 500 cycles | 0.01 M H$_2$SO$_4$ | 13 |
| Cu/Cu$_2$O foam with Ni nanoparticles | 50 | 290 | 100 cycles | 0.1 M H$_2$SO$_4$ | 14 |
| Hierarchical CuO nanoflowers | 67.4 | 1020 | 50 cycles | 0.5 M H$_2$SO$_4$ | 15 |
| Porous Cu2O nanorods | 106 | 184 | 20 hr | 1.0M KOH | 16 |

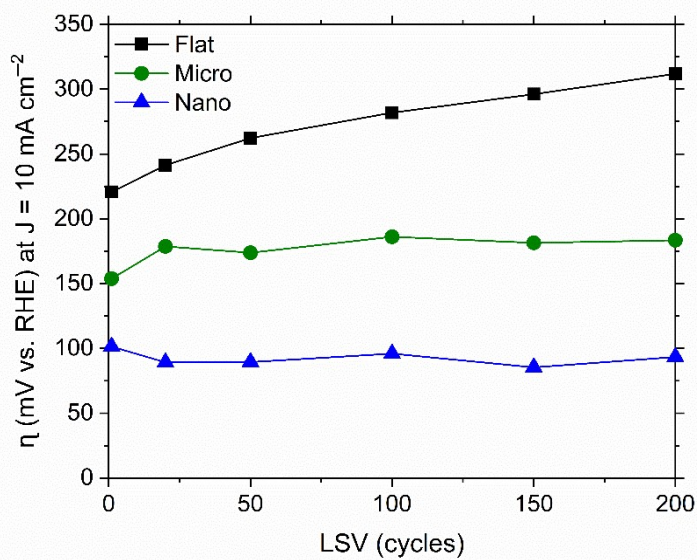

**Figure S1:** Additional stability tests comparing the overpotential η (mV vs. RHE) at J = 10 mA cm$^{-2}$ of flat, micro-patterned and nano-patterned samples up to 200 LSV cycles.

**Additional stability test (experimental method):**

For additional electrochemical measurements, a Solartron XM ModuLab potentiostat was used in combination with a three-electrode cell. The samples were used as working electrode, and silver chloride (Ag|AgCl (E(Ag|AgCl) = 0.197 V vs. SHE) was used as reference electrode. Pt wire was used as counter electrode to complete the three-electrode cell setup. The samples were submerged in 0.5 M H$_2$SO$_4$ solution, and the open circuit potential (OCP) was monitored until it reached a steady state. Subsequently, linear sweep voltammetry (LSV) was performed from -1 V (vs. Ag|AgCl) to 0 V (vs. Ag|AgCl) with a scan rate of 0.02 V/s for 200 cycles on fresh flat, micro-patterned, and nano-patterned samples.

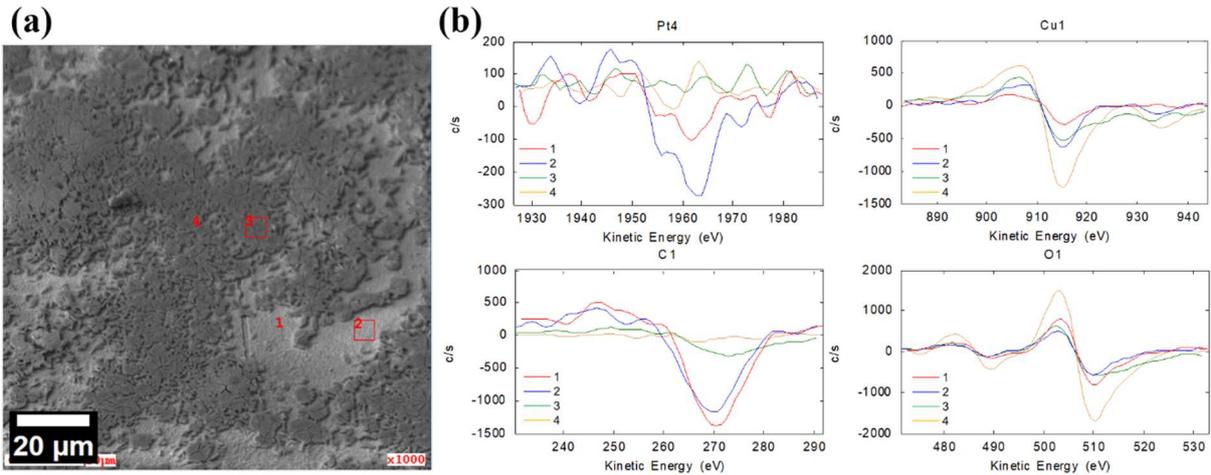

**Figure S2:** AES analysis of the Nano.1000cycle sample. (a) The representative SEM image with markers of selected area and (b) AES spectra of Pt, Cu, C, and O.

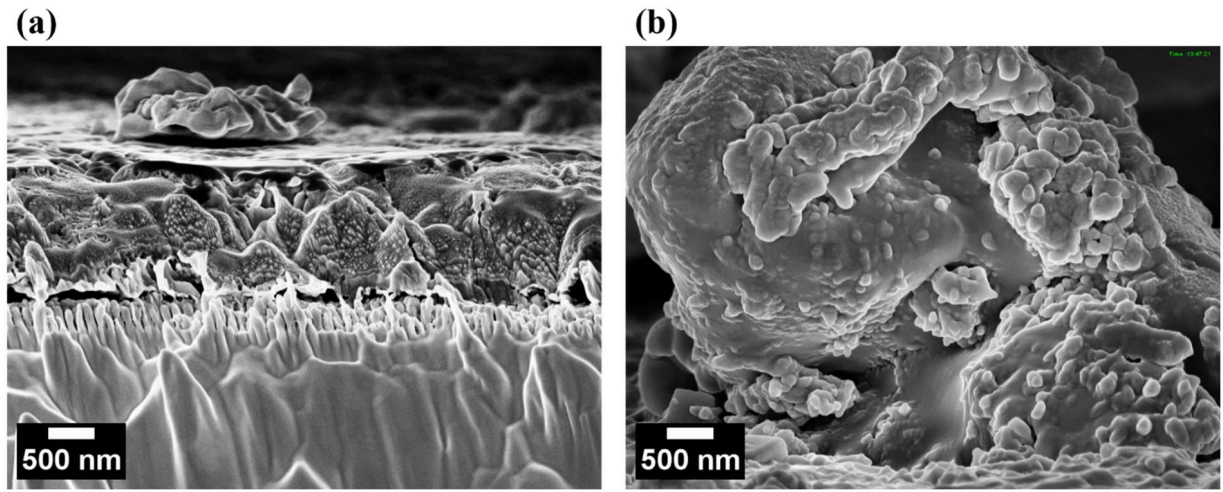

**Figure S3:** SEM images of agglomerated CuO particles on top of the deposited CuO layer. (a) A CuO particle is deposited on the CuO layer. (b) A detailed image of the CuO particle.

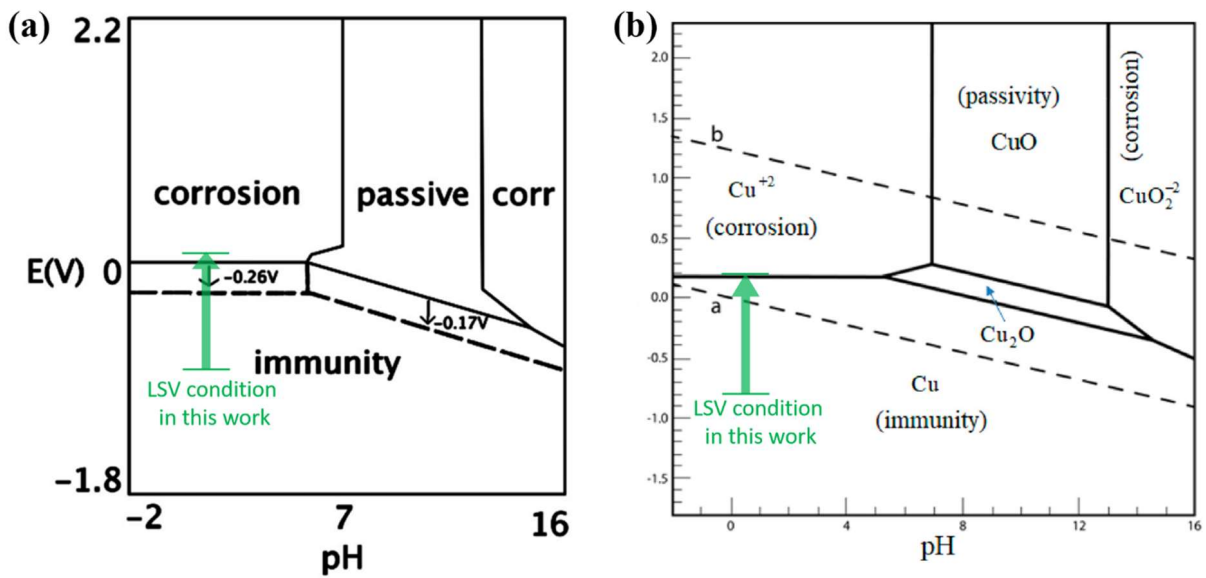

**Figure S4:** Modified Pourbaix diagrams for Cu with green arrows marking the experimental condition. (a) Cu nanomaterials have a negative shift of the dissolution balance potential leading to a reduction in the immunity area

and an expansion of both the passivity and corrosion areas. Reproduced with permission from Taylor et al.[17] (b) A Pourbaix diagram features the oxidation reactions (2Cu + $H_2O$ → $Cu_2O$ + $2H^+$ + $2e^-$) at electrode potentials above the "a" line for hydrogen evolution. Reproduced with permission from Hamidah et al.[18]